\documentclass[12pt]{article}
\usepackage{amsmath,bbm,amssymb,amsfonts}
\usepackage[pdftex]{graphicx}
\newcommand\beq{\begin{equation}}
\newcommand\eeq{\end{equation}}
\newcommand\bea{\begin{eqnarray}}
\newcommand\eea{\end{eqnarray}}

\begin{document}
\vspace{-2.0cm}
\bigskip

\begin{center} 
{\Large \bf Invariants, Projection Operators and
}
\end{center} 
\begin{center} 
{\Large \bf  SU(N)$ \times $SU(N) Irreducible Schwinger Bosons} 
\end{center} 
\vskip .8 true cm

\begin{center} 
{\bf Manu Mathur} \footnote{manu@boson.bose.res.in}, 
{\bf Indrakshi Raychowdhury} \footnote{indrakshi@bose.res.in},{\bf T P Sreeraj} \footnote{sreerajtp@bose.res.in} 
\vskip 1.0 true cm

S. N. Bose National Centre for Basic Sciences \\ 
JD Block, Sector III, Salt Lake City, Calcutta 700098, India

%
%
\end{center} 
\bigskip

\centerline{\bf Abstract}

\noindent We exploit $SU(N)$ Schwinger bosons to construct and analyze the 
coupled irreducible representations of $SU(N) \times SU(N)$ 
in terms of the invariant group.  The corresponding projection operators  
are constructed in terms of the invariant group generators.
We also  construct $SU(N) \times SU(N)$
irreducible Schwinger bosons which directly create these  coupled irreducible 
states.  The $SU(N)$ Clebsch Gordan coefficients 
are computed as the matrix elements of the projection operators.  

\vskip .4 true cm

\section{\bf Introduction}
\label{intros} 

It is well known that the Schwinger representation of the $SU(2)$ Lie algebra
\cite{sch} has played important roles in widely different branches of 
physics such as nuclear physics \cite{np}, condensed matter physics \cite{cmp}, 
quantum optics \cite{qo}, gauge theories \cite{gt}, quantum gravity \cite{qg} etc. 
They have also played equally significant role in the study of  Lie groups 
\cite{perbook}.  In particular, in the context of representation theory of $SU(N)$ group 
\cite{apw} the Schwinger bosons enable us to construct their unitary irreducible representations 
with enormous ease and simplicity \cite{mosh,mmds,nm1,othersb,su3isb,sunisb}. 
The $SU(2)$ case, studied extensively by Schwinger himself, provides the easiest example of  
this simplification. The Hilbert space ${\cal H}_{sb}$ associated with two Schwinger bosons is isomorphic to the 
representation space of $SU(2)$ and is the simplest possible representation or model space 
\cite{gel} of $SU(2)$. 
However, in the case of higher N $(N \ge 3)$, this simple isomorphism 
is lost due to the existence of certain $SU(N)$ invariant operators. These invariant operators follow 
$U(N-1)$ algebra \cite{mosh,nm1,su3isb,sunisb} and lead to $SU(N)$ invariant directions in the Hilbert space 
${\cal H}_{sb}$ associated with $SU(N)$ group. 
Any two states which differ by an overall presence of such invariant operators  will transform in the 
same way under $SU(N)$. 
This leads to the problem of multiplicity which in turn 
makes the representation theory of $SU(N)$ ($N\ge 3$) much  more involved compared 
to $SU(2)$ \cite{mosh,mmds,nm1,su3isb,sunisb}.  The standard way to handle this problem is by  demanding that 
the Schwinger boson states follow the symmetries of $SU(N)$ Young tableaues.  
In \cite{su3isb,sunisb} we showed that these $SU(N)$ Young tableaue symmetries can be easily 
realized by imposing certain $SU(N)$ invariant constraints on  ${\cal H}_{sb}$. 
We further defined $SU(N)$ irreducible Schwinger bosons which weakly commute with the above constraints 
and hence directly create states which are invariant under all $SU(N)$ Young tableaue symmetries 
(see section {\ref{su2isbs}). 
These $SU(N)$ irreducible states created by the monomials of $SU(N)$ irreducible Schwinger bosons are 
also multiplicity free. 
This makes the construction of all $SU(N)$ irreducible representations exactly analogous to the simple $SU(2)$ case. 
Thus the $SU(N)$ invariant constraint formulation provides a novel approach to study the $SU(N)$ representation 
theory. The purpose and motivation of  the present work is to show that the above ideas can also be naturally 
extended to the study  of the coupled representations of the  the 
direct product group $SU(N) \times SU(N)$.  We discuss the simplest and well studied 
$SU(2)\times SU(2)$  case first and then go to higher 
$SU(N)\times SU(N)$  groups. Infact, to our knowledge even the $N=2$  results (section \ref{su2invs}) 
are new and have many novel features.  In particular, we  show that all coupled angular momentum irreducible 
representations can be projected out directly from the decoupled angular momentum states by  certain projection 
operators. These projection operators are built from the invariant $Sp(2,R) \times SU(2)$ 
generators which commute with the total angular momentum generators. The $SU(2)$ Clebsch Gordan coefficients are 
simply the matrix elements of the above projection operators 
in the decoupled basis (see eqn. (\ref{cgc1})) and can be  easily computed (see section \ref{su2cgcs}). 
Further, using the invariant algebra we also construct  $SU(2) \times SU(2)$ irreducible Schwinger bosons which directly 
create all possible coupled $SU(2) \times SU(2)$  irreducible states.  As expected, these simple $SU(2)$ techniques 
based on the invariant groups have natural extention  to all higher $SU(N)$ groups. 
This is significant and important as, inspite of vast amount of literature on $SU(2)$ group and very 
specific techniques valid for $N=3,4$ etc., general computational methods to handle $SU(N)$ group for arbitrary N 
are hard to find \cite{apw,cgc}. 

The plan of the paper is as follows. The section 2 deals with the simplest $SU(2)$ case. 
In this section we briefly construct the total angular momentum group invariant $Sp(2,R) \times 
SU(2)$ algebras \cite{sch}. 
Using this invariant algebra  we construct projection operators which directly project the direct product Hilbert 
space to various irreducible Hilbert spaces characterized by the net angular momentum and net magnetic 
quantum numbers. 
Further, we construct $SU(2) \times SU(2)$ irreducible Schwinger bosons which trivially satisfy the above 
constraints and directly create the direct product irreducible representations. 
We then show that the Clebsch Gordan coefficients can be very easily computed as the matrix elements of 
the projection operators using the invariant algebra. 
In section 3 we show that the above $SU(2)$ techniques have very natural extention to all higher $SU(N)$ groups. 

\begin{center} 
\bf {Direct product representations and invariant algebras} 
\end{center} 

\noindent In the following sections 
we show that the symmetries of the direct product Young tableaues (see Figure \ref{su2f} and Figure \ref{sunf})
can also be realized through certain group invariant constraints  
(see eqn. (\ref{con1})). Further, these group invariant constraints 
lead to projection operators which  project out the coupled irreducible representations 
from the direct product of two irreducible representations. 
As mentioned earlier, we  discuss  the simple  $SU(2)$ group (section \ref{su2invs}) first  
and then generalize these ideas and techniques to $SU(N)$ group with arbitrary N (section \ref{suninvs}). 

\section {Representations of $SU(2) \times SU(2)$ and invariants}
\label{su2invs}

The Schwinger boson representations of $SU(2) \times SU(2)$ Lie algebra is: 
\beq
J^{\mathrm a}_1 ~\equiv ~\frac{1}{2} ~a^{\dagger}_\alpha ~(\sigma^{\mathrm a} )_{\alpha\beta} ~a_\beta, ~~~~~~
J^{\mathrm a}_2 ~\equiv ~\frac{1}{2} ~b^{\dagger}_\alpha ~(\sigma^{\mathrm a} )_{\alpha\beta} ~b_\beta, 
\label{sch} 
\eeq
where $\sigma^{\mathrm a}$ denote the Pauli matrices, 
$(a_\alpha,a^\dagger_\alpha)$ and $(b_\alpha,b^\dagger_\alpha)$ with $\alpha=1,2$ are the 
two Schwinger boson doublets.  It is easy to check that 
the operators in (\ref{sch}) satisfy the $SU(2) \times SU(2)$ Lie algebra with the SU(2) 
Casimirs:  
\bea 
\vec{J}_1 \cdot \vec{J}_1 \equiv {\hat n_a \over 2}  \left({\hat n_a \over 2} + 1\right),  
~~~~~~~~~\vec{J}_2 \cdot \vec{J}_2 \equiv {\hat n_b \over 2}  \left({\hat n_b \over 2} + 1\right).  
\label{casimirs} 
\eea 
In (\ref{casimirs}), $\hat n_a= {\vec{a}^{\dagger} \cdot \vec{a}}=(a^\dagger_1a_1+a^\dagger_2a_2)$ and   
$\hat n_b= {\vec{b}^{\dagger} \cdot \vec{b}} =(b^\dagger_1b_1+b^\dagger_2b_2)$ are the number operators with eigenvalues $n_a=n_a^1+n_a^2$ and $n_b=n_b^1+n_b^2$ respectively. 
The decoupled angular momentum states are: 
\bea 
|j_1,m_1\rangle = \frac{\left(a^\dagger_1\right)^{j_1+m_1} \left(a^\dagger_2\right)^{j_1-m_1}
}{\sqrt{(j_1+m_1)!(j_1-m_1)!}}|0\rangle, ~~~~~~~~ 
|j_2,m_2\rangle = \frac{\left(b^\dagger_1\right)^{j_2+m_2} \left(b^\dagger_2\right)^{j_2-m_2}
}{\sqrt{(j_2+m_2)!(j_2-m_2)!}}|0\rangle. 
\label{dams} 
\eea 
The representations of $SU(2)$ 
can also be characterized by the eigenvalues of the total occupation number operator as,
\bea 
|n^1_a,n^2_a\rangle = \frac{\left(a^\dagger_1\right)^{n_a^1} \left(a^\dagger_2\right)^{n_a^2}
}{\sqrt{n_a^1!n_a^2!}}|0\rangle, ~~ ~~~~
|n_b^1,n_b^2\rangle = \frac{\left(b^\dagger_1\right)^{n_b^1} \left(b^\dagger_2\right)^{n_b^2}
}{\sqrt{n_b^1!n_b^2!}}|0\rangle. 
\label{damso} 
\eea 
In (\ref{damso}) $n_a^1=j_1+m_1, n_a^2=j_1-m_1, n_b^1=j_2+m_2, n_b^2=j_2-m_2.$
The direct product states $|n^1_a,n^2_a\rangle \otimes |n_b^1,n_b^2\rangle$ will often be denoted 
by $\left\vert \begin{array}{cc} n^1_a & n^2_a \\ n^1_b & n^2_b \end{array}\right\rangle.$ 
 The total angular momentum generators are: 
\bea 
J^{\mathrm a}_T = J^{\mathrm a}_1 + J^{\mathrm a}_2. 
\label{su2tsu2}
\eea 
The corresponding group will be denoted by $SU(2)_T$. 
We now construct all possible $SU(2)_T$ invariants out of the two Schwinger boson doublets 
in (\ref{sch}).  The first set of invariant operators is: 
\bea 
k_{+} \equiv a^{\dagger} \cdot \tilde{b}^{\dagger}, ~~k_{-} \equiv a \cdot \tilde{b}, ~~k_0 = 
\frac{1}{2}(\hat n_a+\hat n_b+2).   
\label{sp2r1o} 
\eea 
In (\ref{sp2r1o}) the invariants $k_{\pm}$ are the antisymmetric combination of the two doublets: 
$a^{\dagger} \cdot \tilde{b}^{\dagger} \equiv \epsilon_{\alpha\beta} a^{\dagger}_\alpha b^\dagger_\beta
= (a^\dagger_1b^\dagger_2 - a^\dagger_2b^\dagger_1)$ 
and  $a \cdot \tilde{b} \equiv \epsilon_{\alpha\beta} a_\alpha b_\beta 
= (a_1b_2 - a_2b_1).$ 
It is easy to check that $k_+, k_-$  and $k_0$ commute with $SU(2)_T$   generators 
$J^{\mathrm a}$ in (\ref{su2tsu2}) and satisfy Sp(2,R) algebra: 
\bea 
\left[k_-,k_+\right] = 2k_0, ~~~~~~\left[k_0,k_{\pm}\right] = \pm k_{\pm}.
\label{sp2r1} 
\eea  
The discrete unitary irreducible representations $|k,q\rangle$ of Sp(2,R) relevant for us in this work 
are characterized by the eigenvalues of $k^2\equiv k_1^2+k_2^2-k_0^2 =\frac{1}{2}(k_+k_-+k_-k_+) - k_0^2$ 
and $q$ satisfying: 
\bea 
k_0~|k,q\rangle = q ~|k,q\rangle, ~~~~~~~~k^2~ |k,q\rangle  = k(1-k)~|k,q\rangle.
\label{sp2rev}
\eea 
In (\ref{sp2rev}), $q=k,k+1,k+2,\cdots$. 
The Sp(2,R) raising and lowering operators satisfy: $k_{\pm} |k,q\rangle = \left[(q\pm k)(q\mp k 
\pm 1)\right]^{\frac{1}{2}} |k,q \pm 1\rangle$  and  $k_-|k,q=k\rangle =0$.

\noindent Similarly, another $SU(2)_T$ invariant algebra is obtained by 
defining \cite{sch}: 
\bea 
{\kappa}_{+} \equiv a^{\dagger} \cdot {b}, ~~{\kappa}_-  \equiv b^\dagger \cdot {a}, ~~
\kappa_0 \equiv \frac{1}{2}\left(\hat n_a-\hat n_b\right).   
\label{su2} 
\eea 
These generators satisfy the standard $SU(2)$ algebra: 
\bea 
\left[{\kappa}_+,{\kappa}_-\right] = 2\kappa_0, ~~~\left[\kappa_0,\kappa_{\pm}\right] = \pm \kappa_{\pm}.  
\label{sp2r1u} 
\eea  
It is easy to check that the $Sp(2,R)$ and  $SU(2)$ generators in 
(\ref{sp2r1o}) and (\ref{su2}) respectively commute with each other as well as with $SU(2)_T$ in (\ref{su2tsu2}). 
Therefore, the coupled irreducible representations of $SU(2) \times SU(2)$ can also be labeled by the quantum numbers 
of the $Sp(2,R) \times SU(2)$ group (see (\ref{sp2rc}) and (\ref{su2c})).  \\

\subsection{The Projection operators, invariants and symmetries of Young tableaues} 
\label{su2ps} 

In this section we consider  the coupled angular momentum states obtained by taking the direct product of 
two arbitrary angular momentum states $|j_1,m_1\rangle$ and  $|j_2,m_2\rangle$ as shown in Figure \ref{su2f}. 
\begin{figure}[t]
\centering
\includegraphics[width=17cm,height=5cm]{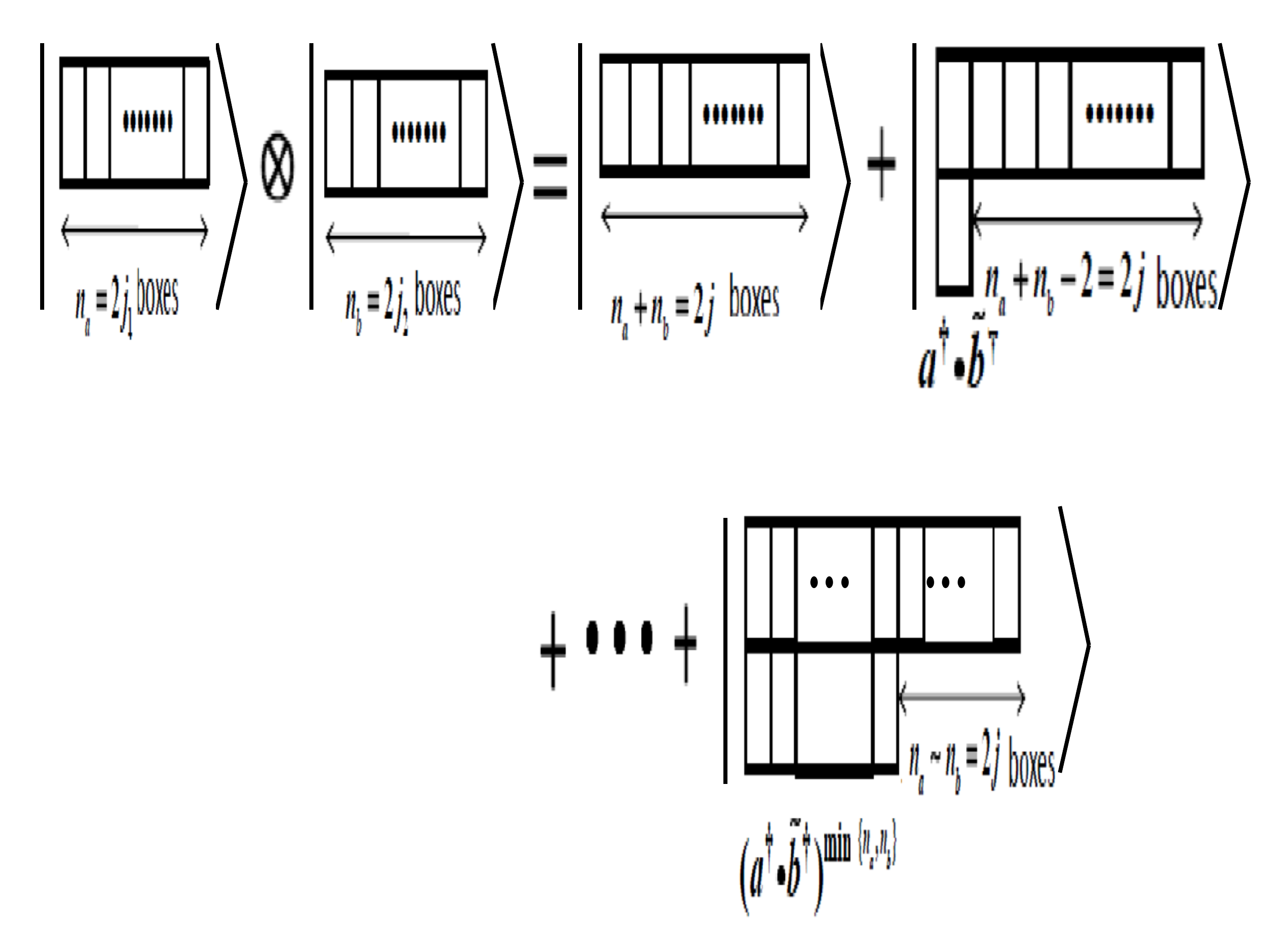}
\caption{Graphical or Young tableaue representation of the identity (\ref{cgd}). The $SU(2) \otimes SU(2)$ 
coupled states on the right hand side  can be directly obtained from the decoupled state by the corresponding 
projection operators (see (\ref{pocgc})).  The coupled states on the right hand sides  also carry   
$Sp(2,R) \times SU(2)$ quantum numbers (see (\ref{sp2rc}) and (\ref{su2c})).} 
\label{su2f} 
\end{figure}
The Young tableaue decomposition in Figure \ref{su2f} corresponds to the standard expansion of the decoupled 
basis in terms of the coupled basis: 
\bea 
|j_1,m_1\rangle \otimes |j_2,m_2\rangle  =  \sum_{j=j_1+j_2}^{|j_1-j_2|} 
C_{j_1,m_1;j_2,m_2}^{~j,m} |j_1,j_2;j,m\rangle,
\label{cgce} 
\eea  
In (\ref{cgce}) $m = m_1+m_2$. 
The same series can also be obtained by defining  projection operators $\mathcal P^j$ which directly project 
the decoupled state to a particular coupled state $|j_1, j_2; j, m\rangle$. In terms of projection 
operators, the expansion (\ref{cgce}) takes the form: 
\bea 
|j_1,m_1\rangle \otimes |j_2,m_2\rangle  =  
\sum_{j=j_1+j_2}^{|j_1-j_2|} {\cal P}^j |j_1,m_1\rangle \otimes |j_2,m_2\rangle 
\equiv  \sum_{r=0}^{min(2j_1,2j_2)} {\cal P}_r 
|j_1,m_1\rangle \otimes |j_2,m_2\rangle. 
\label{cgd} 
\eea 
In (\ref{cgd}) $r$ is the number of two boxes (invariants) on the right hand side of Figure \ref{su2f}, i.e.,  
\bea 
r = j_1+j_2-j. 
\label{ntb} 
\eea 
Comparing the series (\ref{cgd}) with the standard expansion in (\ref{cgce}) 
we get: 
\bea 
{\cal P}_r |j_1,m_1\rangle \otimes |j_2,m_2\rangle  = C_{j_1,m_1;j_2,m_2}^{~j,m} |j_1,j_2;j,m\rangle. 
\label{pocgc} 
\eea 
In (\ref{pocgc}) $j=j_1+j_2-r$ and $m=m_1+m_2$.  Taking the norms of each side of (\ref{pocgc}) and using 
${\cal P}_r^2 ={\cal P}_r$ we get: 
\bea 
C_{j_1,m_1;j_2,m_2}^{~j,m}  = \sqrt{\langle j_1,j_2,m_1,m_2| {\cal P}_r |j_1,j_2,m_1,m_2 \rangle}.  
\label{cgc1} 
\eea 
In (\ref{cgc1}) we have used the notation $|j_1,j_2,m_1,m_2\rangle \equiv |j_1,m_1\rangle \otimes |j_2,m_2\rangle$. 
The Clebsch Gordan coefficients in (\ref{cgc1}) will be explicitly computed in section \ref{su2cgcs}. 

We now construct the projection operators defined in (\ref{cgd}). The Figure {\ref{su2f} and 
(\ref{cgd}) imply that the projection operators can only 
depend on the $SU(2)_T$ invariant operators $\left(\hat n_a, \hat n_b,k_{\pm},{\kappa}_{\pm}\right)$ 
discussed in section 2. We first consider $r=0~ (j=j_1+j_2)$ case. The Figure \ref{su2f} 
implies that ${\cal P}_0 (\equiv {\cal P})$ should completely symmetrize the $SU(2)$ indices so that  $j=j_1+j_2.$ 
Therefore, we demand: 
\bea 
k_- \left({\cal P} |j_1,m_1\rangle \otimes |j_2,m_2 \rangle\right) = 0.   
\label{con1} 
\eea 
As ${\cal P}$ depends only on the $SU(2)_T$ invariant operators, 
we can write the most general form as: 
\bea 
{\cal P} 
= \sum_{q_1,q_2,q_3,q_4}  l_{\{q\}} 
(\hat n_a,\hat n_b) 
(k_+)^{q_1}(k_-)^{q_2} ({\kappa}_+)^{q_3}({\kappa}_-)^{q_4}  
\label{projo} 
\eea 
In (\ref{projo}) $l_{\{q\}} \equiv l_{q_1,q_2,q_3,q_4}(\hat n_a,\hat n_b)$ are the number operator dependent 
operators. Further, as the projection operator should not change the number of either a or b type oscillators, 
we get $q_1 = q_2, ~q_3=q_4.$  On the other hand, the identity $\epsilon_{\alpha\beta}\epsilon_{\gamma\delta} 
= \delta_{\alpha\gamma}\delta_{\beta\delta} - \delta_{\alpha\delta}\delta_{\beta\gamma}$ implies:  
\bea 
{\kappa}_+ {\kappa}_- = \hat n_a\hat n_b -k_+k_-.
\label{rel} 
\eea 
Thus all $SU(2)$ operators ${\kappa}_+{\kappa}_-$ in (\ref{projo}) can be removed in terms of Sp(2,R) 
operators $k_+k_-$. 
Therefore, the most general form of the projection operator is: 
\bea 
{\cal P} 
= \sum_{q=0}^{\infty}  l_q(\hat n_a,\hat n_b)~ (k_+)^{q}(k_-)^{q}. 
\label{mgpo}
\eea 
The constants  $l_q$ can be computed (see appendix A.1) by using the constraint (\ref{con1}) and lead to: 
\bea  
l_{q}(\hat n_a,\hat n_b)= \frac{(-1)^q}{q!} 
\frac{(\hat n_a+\hat n_b-q)!}{(\hat n_a+\hat n_b)!} 
\label{pexp} 
\eea 
Note that the constant term in (\ref{pexp}) is chosen to be unity (i.e $l_0=1$) so that: 
$${\cal P}^2 = {\cal P}{\cal P} = \left(1 + \sum_{q=1}^{\infty}  
l_q(\hat n_a,\hat n_b)~ (k_+)^{q}(k_-)^{q}\right) {\cal P}   = 
{\cal P}$$ 
as $k_-{\cal P} =0.$  
The Figure \ref{su2f} now immediately implies that all other projection operators are of the 
form: 
\bea 
{\cal P}_r = N_r (k_+)^r~ {\cal P}~ (k_-)^r = N_r (k_+)^r~ {\cal P}~ (k_-)^r 
\label{rpr} 
\eea 
The constant coefficients $N_r$ are fixed by demanding that the operators ${\cal P}_r$ satisfy 
${\cal P}_r^2 ={\cal P}_r$ (see appendix A.1).  
We thus get: 
\bea 
N_r = \frac{(n_a+ n_b-2r+1)!}{r!( n_a+ n_b-r+1)!} 
\label{conp}
\eea 
Note that these coefficient can also be computed by demanding completeness property: 
\bea
\sum_{r=0}^{\textrm{min}(2j_1,2j_2)} {\cal P}_r = {\it I}. 
\label{idsu2}
\eea
The completeness property (\ref{idsu2})  which is manifest in the defining  expansion  (\ref{cgd}) 
is proved in appendix A.1.
It is also easy to check that the Hilbert spaces projected by different projection operators in 
(\ref{rpr}) are orthogonal: 
\bea 
{\cal P}_r {\cal P}_s = \delta_{rs} {\cal P}_r, 
~~~ r,s = 0,1,2, \cdots \textrm{ min} (2j_1,2j_2). 
\label{ido} 
\eea 
In (\ref{ido}) we have used the Sp(2,R) commutation relation (\ref{sp2r1o}) and the constraints 
$k_-{\cal P}= 0~ (r > s),  ~~ {\cal P} k_+ = 0 ( r <  s).$
We note that the coupled angular momentum states ${\cal P} k_-^r  
|j_1,m_1\rangle \otimes |j_2,m_2\rangle$ 
in the expansion (\ref{cgd}) also belong to the Sp(2,R) irreducible representations with 
lowest Sp(2,R) magnetic quantum number $q=q_0 = (j_1+j_2-r+1)$ as: 
\bea 
k_0 ~{\cal P}~ \left(k_-\right)^r |j_1,m_1\rangle \otimes |j_2,m_2\rangle   & = & 
q_0 ~{\cal P}  \left(k_-\right)^r |j_1,m_1\rangle \otimes |j_2,m_2\rangle . \nonumber \\ &&  \label{abc1} \\ 
k^2 ~{\cal P} \left( k_-\right)^r |j_1,m_1\rangle \otimes |j_2,m_2\rangle &  = & 
q_0(1-q_0) ~{\cal P} \left( k_-\right)^r |j_1,m_1\rangle \otimes |j_2,m_2\rangle. \nonumber 
\eea  
To get the second eigenvalue equation we have used $k_-{\cal P} =0$ 
to replace $k^2 (\equiv \frac{1}{2}(k_+k_-+k_-k_+) - k_0^2)$ by 
$\frac{1}{2}[k_-,k_+] - k_0^2 = k_0(1-k_0).$  The equations (\ref{abc1}) immediately imply: 
\bea 
k_0 ~{\cal P}_r~ |j_1,m_1\rangle \otimes |j_2,m_2\rangle   & = & 
(j_1+j_2+1)~{\cal P}_r |j_1,m_1\rangle \otimes |j_2,m_2\rangle . \nonumber \\    
\label{sp2rc} \\
k^2 ~{\cal P}_r |j_1,m_1\rangle \otimes |j_2,m_2\rangle &  = & 
q_0(1-q_0) ~{\cal P}_r |j_1,m_1\rangle \otimes |j_2,m_2\rangle. \nonumber  
\eea  
Similarly, it is easy to check that the quantum numbers of $SU(2)_T$ invariant $SU(2)$ group in 
(\ref{su2}) are: 
\bea 
\kappa_0 ~{\cal P}_r~ |j_1,m_1\rangle \otimes |j_2,m_2\rangle   & = & 
(j_1-j_2)~{\cal P}_r |j_1,m_1\rangle \otimes |j_2,m_2\rangle, \nonumber \\    
\label{su2c} \\
\kappa^2 ~{\cal P}_r |j_1,m_1\rangle \otimes |j_2,m_2\rangle &  = & 
j(j+1) ~{\cal P}_r |j_1,m_1\rangle \otimes |j_2,m_2\rangle. \nonumber  
\eea  
Note that $j=j_1+j_2-r$ in (\ref{su2c}).

\subsection{$SU(2) \times SU(2)$ irreducible Schwinger bosons} 
\label{su2isbs} 

It is known that all possible $SU(N)$ irreducible representations can be written as 
monomials of $SU(N)$ irreducible Schwinger bosons \cite{su3isb,sunisb}.  This 
construction is the $SU(N)$ extension of the Schwinger $SU(2)$ construction \cite{sch}.  
In this section we  apply these  ideas to construct the coupled states $|j_1,j_2,j,m\rangle$  
in (\ref{pocgc})  as monomials of $SU(2) \times SU(2)$ irreducible Schwinger bosons (see equation (\ref{fie})). 
The $SU(2) \times SU(2)$ irreducible Schwinger boson creation 
operators create states which satisfy $k_-=0$ and therefore correspond to maximally symmetric 
(or states with highest angular momentum) states. All other states can be constructed  by 
applying the  invariant operators on such  maximally symmetric states.  Note that this procedure 
is also illustrated by Figure \ref{su2f}. The first coupled state on the 
right hand side with $n_a+n_b =2j_1+2j_2$ is the maximally symmetric state. All other coupled states 
on the right hand side are obtained by multiplications of the invariant $k_+$ (i.e., two boxes arranged 
vertically in Figure \ref{su2f}) on such  maximally symmetric states.      
As in \cite{su3isb,sunisb}, we define: 
\bea 
A^\dagger_\alpha \equiv a^\dagger_\alpha + f(\hat n_a,\hat n_b) k_+ \tilde{b}_\alpha,  ~~ B^\dagger_\alpha \equiv 
b^\dagger_\alpha + g(\hat n_a,\hat n_b) k_+ \tilde{a}_\alpha. 
\label{dpisb}
\eea   
Note that by construction (\ref{dpisb})  the $SU(2) \times SU(2)$ transformation properties  of 
$A^\dagger_\alpha$  and $B^\dagger_\alpha$  are exactly same as those of $a^\dagger_\alpha$  
and $b^\dagger_\alpha$ respectively.  We now demand: 
\bea   
k_- ~A^\dagger_\alpha ~{\cal P} |j_1,m_1\rangle \otimes |j_2,m_2\rangle = 0, ~~~ 
k_-~ B^\dagger_\alpha ~{\cal P} |j_1,m_1\rangle \otimes |j_2,m_2\rangle = 0. 
\label{conab} 
\eea 
The above constraints can be solved in terms of the unknown operator valued functions  $f(\hat n_a,\hat n_b)$ and 
$g(\hat n_a,\hat n_b)$: 
\bea 
f(\hat n_a,\hat n_b) = - \frac{1}{(\hat n_a+\hat n_b)}, ~~~~~ g(\hat n_a,\hat n_b) =  \frac{1}{(\hat n_a+\hat n_b)}.  
\label{conf}
\eea 
Note that the $f(\hat n^a, \hat n^b)$  and $g(\hat n^a, \hat n^b)$   in (\ref{conf}) are well defined as they 
always follow a creation operator in (\ref{dpisb}). 
As an example of the states created by the $SU(2) \times SU(2)$ irreducible Schwinger bosons we consider the state: 
$A^\dagger_\alpha B^\dagger_\beta |0\rangle =  
A^\dagger_\beta B^\dagger_\alpha |0\rangle = \frac{1}{2} 
\left(a^\dagger_\alpha b^\dagger_\beta + a^\dagger_\beta  b^\dagger_\alpha\right) |0\rangle$.  
We note that it is already symmetric in the $SU(2)$ indices $\alpha$ and $\beta$ and 
no explicit symmetrization is needed. Infact, 
\bea 
A^\dagger_1 B^\dagger_1 |0\rangle & = &  ~~|j_1=1/2,j_2=1/2,j=1,m =1\rangle, \nonumber \\
A^\dagger_1 B^\dagger_2 |0\rangle &  =&  \frac{1}{2}  |j_1=1/2,j_2=1/2,j=1,m =0\rangle,   \\
A^\dagger_2 B^\dagger_2 |0\rangle &  = &  ~~|j_1=1/2,j_2=1/2,j=1,m =-1 \rangle. \nonumber   
\eea 
The irreducible Schwinger bosons can also be directly constructed using the projection operators 
of the previous section as: 
\bea 
A^\dagger_\alpha \approx {\cal P} a^\dagger_\alpha, ~~~~ B^\dagger_\alpha \approx {\cal P} b^\dagger_\alpha. 
\label{rhs} 
\eea  
In (\ref{rhs}) $\approx$ implies weak equality.  In other words the equations 
(\ref{rhs}) are true only on 
the projected section of the Hilbert space which satisfies the constraint $k_-=0$. 
The equivalence of (\ref{rhs}) and (\ref{dpisb}) can be easily established by substituting ${\cal P}$ 
from (\ref{mgpo}) in (\ref{rhs}) and noting that $l_1(\hat n_a, \hat n_b) = 
f(\hat n_a, \hat n_b) = - g(\hat n_a, \hat n_b)$. 
The completely symmetric states of $SU(2)_T$ can be easily defined through 
the irreducible Schwinger bosons: 
\bea 
|j_1,j_2,j=j_1+j_2,m \rangle  & \equiv & {N}^{j_1m_1}_{j_2m_2}~  
\frac{\left(A^\dagger_1\right)^{j_1+m_1}
\left(A^\dagger_2\right)^{j_1-m_1} \left(B^\dagger_1\right)^{j_2+m_2} \left(B^\dagger_2\right)^{j_2-m_2}}
{\sqrt{(j_1+m_1)!(j_1-m_1)!(j_2+m_2)!(j_2-m_2)!}} 
~|0\rangle . 
\label{jm12}
\eea 
To compute the normalization constant ${N}^{j_1,m_1}_{j_2,m_2}$ in (\ref{jm12}) we note that the right 
hand side of the above equation can also be written in terms of decoupled states as: 
\bea 
\left(N_{j_1m_1}^{j_2m_2}\right)^{-1} |j_1,j_2;j=j_1+j_2,m\rangle & = & 
{\cal P}~ \frac{\left(A^\dagger_1\right)^{j_1+m_1}
\left(A^\dagger_2\right)^{j_1-m_1} \left(B^\dagger_1\right)^{j_2+m_2} \left(B^\dagger_2\right)^{j_2-m_2}}
{\sqrt{(j_1+m_1)!(j_1-m_1)!(j_2+m_2)!(j_2-m_2)!}} 
~|0\rangle \nonumber \\  
& = &     
{\cal P}~ \frac{\left(a^\dagger_1\right)^{j_1+m_1}
\left(a^\dagger_2\right)^{j_1-m_1} \left(b^\dagger_1\right)^{j_2+m_2} \left(b^\dagger_2\right)^{j_2-m_2}}
{\sqrt{(j_1+m_1)!(j_1-m_1)!(j_2+m_2)!(j_2-m_2)!}} 
~|0\rangle  \nonumber \\ \nonumber \\
& = & 
{\cal P} |j_1,m_1\rangle \otimes |j_2,m_2\rangle.  
\label{rel2} 
\eea 
In the first step above  we have introduced identity as ${\cal P}$. 
We then replace the irreducible Schwinger bosons by their defining equations (\ref{dpisb}) and  used ${\cal P} k_+ =0$ 
in the second step to get the decoupled states at the end.  To compute the normalization $N^{j_1m_1}_{j_2m_2}$ 
in (\ref{jm12}) we notice that the completely symmetric states are given by (\ref{pocgc}) at $r=0$: 
\bea 
{\cal P} ~|j_1,m_1\rangle \otimes |j_2,m_2\rangle = C^{j=j_1+j_2,m}_{j_1,m_1,j_2,m_2} 
|j_1,j_2;j=j_1+j_2,m\rangle.  \nonumber 
\eea 
Comparing this with (\ref{rel2}) we get: 
${N}^{j_1,m_1}_{j_2,m_2} ~C^{j=j_1+j_2,m}_{j_1,m_1,j_2,m_2} =1$.
Therefore, 
\bea 
{N}^{j_1m_1}_{j_2m_2}  = \frac{1}{C^{j=j_1+j_2,m}_{j_1,m_1,j_2,m_2}} 
= \left[\frac{(2j_1+2j_2)!(j_1+m_1)!(j_1-m_1)!(j_2+m_2)!(j_2-m_2)!}
{(2j_1)!(2j_2)!(j_1+j_2+m_1+m_2)!(j_1+j_2-m_1-m_2)!}\right]^{1/2} 
\label{norm1}
\eea 
For example, we put $j_1=2,m_1=0,j_2=1,m_2=0$ in (\ref{jm12}) and replace the irreducible 
Schwinger bosons by their defining equations (\ref{dpisb}) and (\ref{conf}) to get: 
\bea 
|j_1= 2,j_2=1,j=3,m=0 \rangle = {N}^{20}_{10} \left[\frac{3}{5} |2,0 \rangle |1,0 \rangle + 
\frac{\sqrt{3}}{5} |2 ,-1 \rangle |1,1 \rangle + \frac{\sqrt{3}}{5} |2, 1 \rangle |1,-1 \rangle\right]. 
\label{ccxu} 
\eea 
Therefore,  explicit normalization of the above state gives:  
${N}^{20}_{10} = \sqrt{\frac{5}{3}}$  which is also the value 
obtained by (\ref{norm1}) with 
$C^{~~j=3,m=0}_{j_1 = 2,m_1=0;j_2=1,m_2=0}=\sqrt{\frac{3}{5}}$. 
With this value of normalization, the expansion (\ref{ccxu}) further gives: 
$$C^{~~j=3,m=0}_{j_1 = 2,m_1=-1;j_2=1,m_2=1}= 
C^{~~j=3,m=0}_{j_1 = 2,m_1=1;j_2=1,m_2=-1}= \sqrt{\frac{1}{5}}.$$  
The same values are also obtained from the Clebsch Gordan series (\ref{ffcg}) obtained using 
the invariants in the next section. The above example provides a self consistency check on 
the procedure. 
The discussions in the previous section imply that  an arbitrary coupled state can be written as: 
\bea 
|j_1,j_2;j,m\rangle =  
{\cal N}^{~j}_{j_1,j_2} \left(k_+\right)^{j_1+j_2-j} 
|(j_1-j_2+j)/2,(j_2-j_1+j)/2;j,m\rangle.  
\label{fie}
\eea 
Note that the state $ |(j_1-j_2+j)/2,(j_2-j_1+j)/2;j,m\rangle$ is maximally symmetric and is 
of the form (\ref{jm12}). 
The normalization 
constants ${\cal N}^{~j}_{j_1,j_2}$ can 
be easily computed using 
the commutation relations (\ref{sp2r1}) as $k_- |(j_1-j_2+j)/2,(j_2-j_1+j)/2;j,m\rangle =0$. They are given by: 
$${\cal N}^{~j}_{j_1,j_2} = \sqrt{\frac{(2j_1+2j_2+1)!}{(j_1+j_2-j)!(3j_1+3j_2-j +1)!}}.$$  
We again emphasize that  all possible $SU(2)\times SU(2)$ coupled states in (\ref{fie}) are 
monomials of the irreducible Schwinger bosons. All the symmetries of the coupled Young tableaues 
on the right hand side of Figure \ref{su2f} are already present in (\ref{fie}) and there is no need for 
explicit symmetrization or anti-symmetrization by hand.  
Thus the the irreducible Schwinger bosons (\ref{fie}) can be thought of as the 
generalization of $SU(2)$ Schwinger bosons (\ref{dams}) which directly lead to coupled angular momentum states. 

\noindent The $SU(2) \times SU(2)$ irreducible Schwinger bosons satisfy the following algebra: 
\bea 
\label{ccr2}
\left[A^\dagger_\alpha,A^\dagger_\beta\right] =0,~
\left[B^\dagger_\alpha,B^\dagger_\beta\right] =0,~
\left[A^\dagger_\alpha,B^\dagger_\beta\right] =0. ~~~~~~~~~~~~~~~~~\nonumber \\  
\left[A_\alpha,A^\dagger_\beta\right] = \delta_{\alpha\beta}-\frac{1}{\hat n_a+\hat n_b+1}b^\dagger_\alpha b_\beta+ \frac{1}{(\hat n_a+\hat n_b)(\hat n_a+\hat n_b+1)}k_+ \tilde{b}_\beta a_\alpha  \\
\left[B_\alpha,B^\dagger_\beta\right] = \delta_{\alpha\beta}+\frac{1}{\hat n_a+\hat n_b+1}a^\dagger_\alpha a_\beta - \frac{1}{(\hat n_a+\hat n_b)(\hat n_a+\hat n_b+1)}k_+ \tilde{a}_\beta b_\alpha.  \nonumber  
\eea

\subsection{The projection operators and $SU(2)$ Clebsch Gordan Coefficients} 
\label{su2cgcs} 

The Clebsch Gordan coefficients are given by the defining equation (\ref{pocgc}): 
\bea 
\label{cgc}
C_{j_1,m_1;j_2,m_2}^{j=j_1+j_2-r,m}   =
\langle j_1,j_2;j=j_1+j_2-r,m| {\cal P}_r |j_1,m_1,j_2,m_2\rangle 
\label{des} 
\eea
This can be rewritten as: 
\bea
C_{j_1,m_1;j_2,m_2}^{j=j_1+j_2-r,m} & = & \frac{\langle j_1,m^\prime_1=j_1,j_2,m^\prime_2=m -j_1| 
{\cal P}_r {\cal P}_r|j_1,m_1,j_2,m_2\rangle}  
{ C^{j=j_1+j_2-r,m}_{j_1,m_1^\prime = j_1;j_2,m_2'=m-j_1}} \nonumber \\&=&
\frac{\langle j_1,j_1,j_2,m -j_1| {\cal P}_r |j_1,m_1,j_2,m_2\rangle}  
{\left[\langle j_1,j_1, j_2,m -j_1| {\cal P}_r 
|j_1,j_1,j_2,m-j_1\rangle\right]^{\frac{1}{2}}}  
\label{cgce2} 
\eea  
We have used ${\cal P}_r^2 = {\cal P}_r$ in (\ref{cgce2}). As shown in appendix B.1, the above matrix 
elements of ${\cal P}_r$ can be easily computed 
to give: 
\bea 
C_{j_1,m_1;j_2,m_2}^{~~j,m}&& \hspace{-0.51cm} = \delta_{m, m_1+m_2} 
\sqrt{\frac{(j_1-j_2+j)!(j_2-m_2)!(j_2+m_2)!(j_1+m_1)! (2j+1) (j-m)!}{(j_1+j_2+j+1)!(j_2-j_1+j)!(j_1+j_2-j)!
(j_1-m_1)!(j+m)!}} \nonumber \\ \nonumber \\
&& \hspace{-0.4cm} \sum_{q=0}^{\mbox{min}\{j_1-j_2+j,j_2-j_1+j\}}   \frac{(-1)^
{q+j_1-m_1}(2j-q)!(j_1+j_2-j+q)!}{q!(j_1-j_2+j-q)!(j-m-q)!(j_2-j+m_1+q)!}
\label{ffcg} 
\eea 
The series representing Clebsch Gordon coefficient in (\ref{ffcg}) matches with the expansion given 
in \cite{varsha}.  In section \ref{suncgcs} this $SU(2)$ computation will be extended to $SU(N)$. 

\section{Invariants and representations of $SU(N) \times SU(N)$} 
\label{suninvs} 

We now generalize the previous $SU(2)$ ideas and techniques to  direct product of two conjugate 
representations of $SU(N)$. For simplicity we choose these to be $N$ and $N^*$ 
representations of $SU(N)$. We write the corresponding generators as: 
\beq
Q^{\mathrm a}_1 ~\equiv ~\frac{1}{2} ~{a^{\dagger}}^\alpha ~(\Lambda^{\mathrm a} )_{\alpha}^{~\beta} ~a_\beta,
~~~~Q^{\mathrm a}_2 ~\equiv -~\frac{1}{2} ~b^{\dagger}_\alpha 
~(\tilde{\Lambda}^{\mathrm a} )^{\alpha}_{~\beta} ~b^\beta. 
\label{schn} 
\eeq
In (\ref{schn}) ${\mathrm a}=1,2,\cdots (N^2-1)$ and $\alpha,\beta = 1,2 \cdots N$. $\Lambda^{\mathrm a}$'s are the 
generalized Gell-Mann matrices for $N$-plets of $SU(N)$ and $-\tilde{\Lambda}^{\mathrm a}$ are the dual 
matrices  corresponding to the $N^\ast$-plets of $SU(N)$. From (\ref{schn}) it is clear that $a^\dagger$'s 
transform as $N$ under one $SU(N)$ and $b^\dagger$'s transform as $N^\ast$ under another $SU(N)$\footnote{
For $N \ge 3$ the N and $N^*$ representations are not equivalent. Therefore, we now use upper 
$a^{\dagger \alpha}$ and lower $b^\dagger_\alpha$ indices to differentiate between the two 
conjugate representations.}.  
Like in $SU(2)$ case (\ref{damso}) the decoupled $N$ and $N^*$ irreducible representations 
are:
\bea 
|\{n_a\}\rangle\equiv| n_a^1,n_a^2,\cdots n_a^N\rangle & = & \frac{\left(a^{\dagger 1}\right)^{n_a^1} 
\left(a^{\dagger 2}\right)^{n_a^2} 
\cdots  \left(a^{\dagger N}\right)^{n_a^N}}
{\sqrt{n_a^1! n_a^2! \cdots n_a^N!}} |0\rangle,  \nonumber \\
|\{n_b\}\rangle\equiv|n_b^1,n_b^2,\cdots n_b^N\rangle  & = & \frac{\left(b^\dagger_1\right)^{n_b^1} 
\left(b^\dagger_2\right)^{n_b^2} \cdots  \left(b^\dagger_N\right)^{n_b^N}}
{\sqrt{n_b^1!n_b^2! \cdots n_b^N!}} |0\rangle 
\label{dprn} 
\eea
\begin{figure}[t]
\centering
\includegraphics[width=16cm,height=6cm]{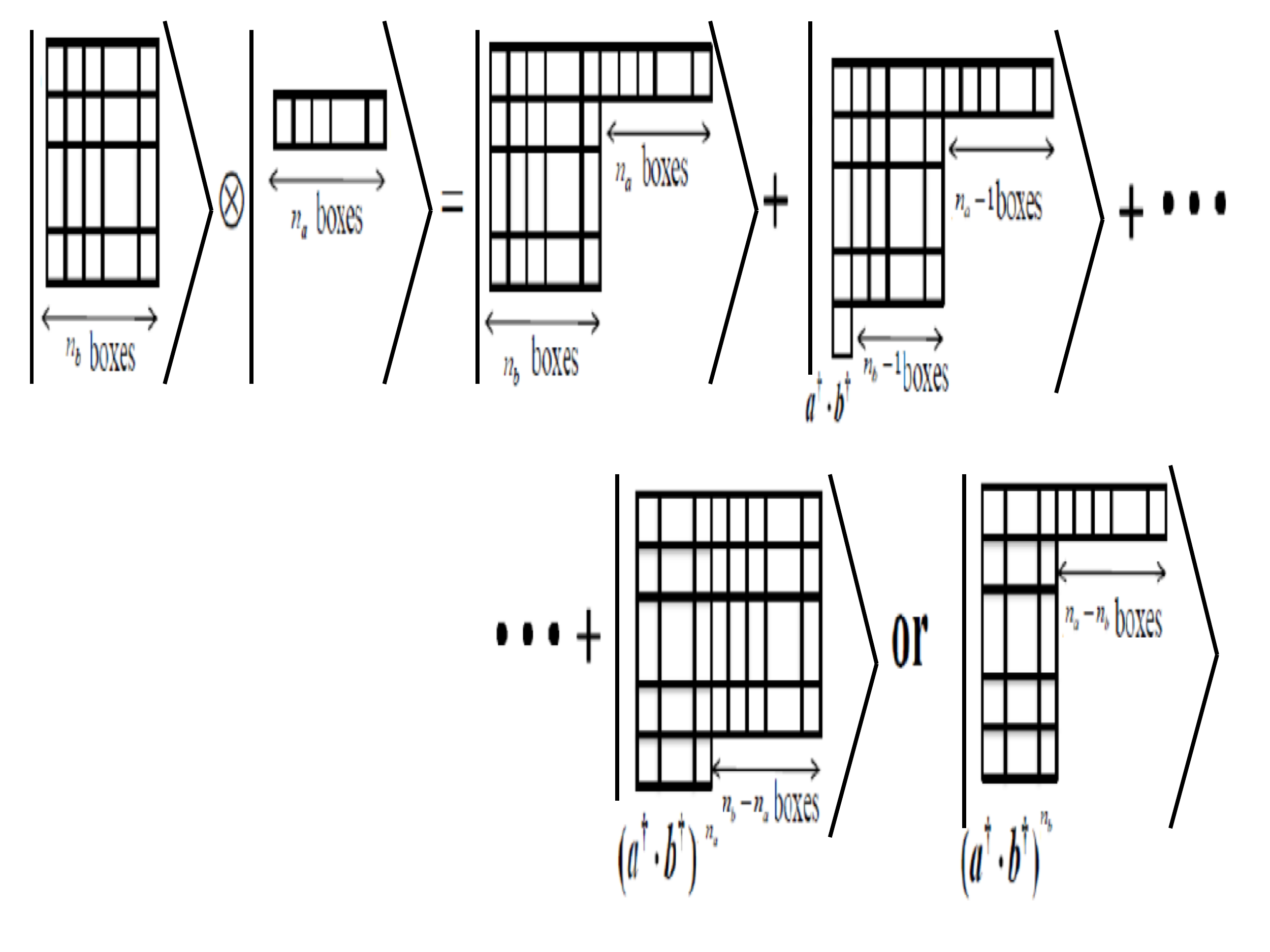}
\caption{Graphical or Young tableau representation of the identity (\ref{suncgd}). The coupled $SU(N) \times 
SU(N)$ states on the right hand side can be directly obtained by projection operators (see (\ref{suncgc1}))
and also carry Sp(2,R) quantum numbers (see (\ref{sp2rnn})).}  
\label{sunf} 
\end{figure}
In (\ref{dprn})  $\{n\}$ represents N partitions of n. 
The two Casimirs are the two total number operators $\hat n_a$ and $\hat n_b$ with eigenvalues $n_a$ and $n_b$ 
respectively: 
\bea 
n_a^1+ n_a^2+ \cdots +n_a^N = n_a, ~~~~~ n_b^1+ n_b^2+ \cdots + n_b^N =n_b.  
\label{casn}
\eea 
We will often denote the $SU(N)$ direct product state 
$|\{n_a\}\rangle \otimes |\{n_b\}\rangle$  by 
$\left\vert \begin{array}{cccc} n_a^1 & n_a^2 & \cdots & n_a^N \\ n_b^1  &n_b^2 &\cdots& n_b^N 
\end{array}\right\rangle.$  As in the case of $SU(2)$ (\ref{su2tsu2}), we define the total $SU(N)$ flux operators:  
\bea 
Q^a_T = Q^a_1+Q^a_2. 
\label{tosun}
\eea 
The corresponding group will be denoted by $SU(N)_T$. At this stage we can also define the coupled 
$SU(N)\times SU(N)$ states through the Clebsch Gordan decomposition (see Figure \ref{sunf}) as: 
\bea 
|\{n_a\}\rangle \otimes |\{n_b\}\rangle  = 
\sum_{r=0}^{min(n_a,n_b)} {C}^{~~r}_{\{n_a\},\{n_b\} }~ |\{n_a-r\}\{n_b-r\};r \rangle  
\label{suncgd0} 
\eea 
As in Figure \ref{sunf}, the coupled states denoted by $|\{n_a-r\}\{n_b-r\};r \rangle$ in (\ref{suncgd0})  
represents the invariant operator $(k_+)^r$ acting on the completely traceless tensor state of rank 
$(n_a-r,n_b-r)$. 

\noindent We now define the following  $SU(N) \times SU(N)$ invariant operators: 
\bea 
k_{+} \equiv a^{\dagger} \cdot {b}^{\dagger}, ~~k_{-} \equiv a \cdot {b}, ~~k_0 = \frac{1}{2}(\hat n_a+\hat n_b+N).   
\label{sp2r1on} 
\eea 
In(\ref{sp2r1on}) the invariants are the scalar products of $N$ and $N^*$ representations: 
$a^{\dagger} \cdot {b}^{\dagger} 
= (a^{\dagger 1}b^\dagger_1 + a^{\dagger 2}b^\dagger_2 + \cdots +a^{\dagger N} b^\dagger_N)$ and 
$a \cdot {b}  = (a_1b^1 + a_2b^2 + \cdots + a_Nb^N).$ 
It is easy to check that they again satisfy Sp(2,R) algebra (\ref{sp2r1}): 
\bea 
\left[k_-,k_+\right] = 2k_0, ~~~~~~\left[k_0,k_{\pm}\right] = \pm k_{\pm}. 
\label{sp2r11} 
\eea  
Like in $SU(2)$ case, the $SU(N) \times SU(N)$ projection operators are defined as: 
\bea 
|\{n_a\}\rangle \otimes |\{n_b\}\rangle  = 
\sum_{r=0}^{min(n_a,n_b)} {\cal P}_r~ |\{n_a\}\rangle \otimes |\{n_b\}\rangle.   
\label{suncgd} 
\eea 
Comparing  (\ref{suncgd0}) with (\ref{suncgd}) we get: 
\bea 
{\cal P}_r |\{n_a\}\rangle \otimes |\{n_b\}\rangle \equiv C^{~~~r}_{\{n_a\},\{n_b\}} |\{n_a-r\},\{n_b-r\};r\rangle.  
\label{suncgc1}
\eea 
In (\ref{suncgc1})  $ C^{~~~r}_{\{n_a\},\{n_b\}}$ are the Clebsch Gordan coefficients. Taking the norms 
on both the sides of (\ref{suncgc1}) we get a simple expression for the $SU(N)$ Clebsch Gordan coefficients: 
\bea 
C^{~~~r}_{\{n_a\},\{n_b\}} = 
\sqrt{\langle \{n_a\}| \otimes \langle \{n_b\} |~{\cal P}_r~|\{n_a\}\rangle \otimes |\{n_b\}\rangle}  
\label{secgc}
\eea We will explicitly compute these coefficients in section \ref{suncgcs}. 

\noindent We now construct the projection operators defined in (\ref{suncgc1}). 
Note that  the condition of tracelesness is exactly same as demanding  the constraint $k_-=0$. 
Using this fact and the invariant algebra (\ref{sp2r11}) the $SU(N)$ projection operator ${\cal P}_0$ 
can be easily constructed like in $SU(2)$ case (see appendix A.2):  
\bea 
{\cal P} \equiv {\cal P}_0 = \sum_{q=0}^{\infty} L_q\left(\hat n^a, \hat n^b\right)  (k_+)^q(k_-)^q 
\label{sunpo} 
\eea 
where
\bea 
L_q\left(\hat n_a, \hat n_b\right) = \frac{(-1)^q}{q!}\frac{(\hat n_a+\hat n_b+N-2-q)!}{(\hat n_a 
+\hat n_b+N-2)!}. 
\label{sunpo1})
\eea 
Again, the projection operator  in (\ref{sunpo}) satisfies 
$${\cal P}^2 = {\cal P}{\cal P}=  \left(1 - \frac{1}{\hat n_a+\hat n_b +N-2} k_+k_- + \cdots\right) {\cal P} 
= {\cal P}, $$ 
as $k_-{\cal P} =0$: 
Note that the $SU(N)$ projection operator (\ref{sunpo}) reduces to the $SU(2)$ projection operator 
(\ref{mgpo}) at $N=2$. Like in $SU(2)$ case, all other projection operators in (\ref{suncgd}) 
or equivalently in  Figure \ref{sunf} are of the form: 
\bea 
{\cal P}_r = N_r ~~(k_+)^r~ {\cal P}_0~ (k_-)^r = N_r ~~ (k_+)^r ~ {\cal P}~ (k_-)^r 
\label{sunpoh} 
\eea 
The constant coefficients $N_r$ are fixed by demanding that the operators ${\cal P}_r$ satisfy:  
${\cal P}_r^2 ={\cal P}_r$ and are given by (see appendix A.2): 
\bea 
N_r = \frac{(n_a+ n_b+N-2r-1)!}{r!( n_a+n_b+N-r-1)!} 
\label{conpn}
\eea 
As expected, (\ref{conpn}) reduces to (\ref{conp}) at $N=2$. 
Like in $SU(2)$ case the projection operators satisfy the orthogonality and completeness properties: 
\bea 
\sum_{r=0}^{\mbox{min} (n_a, n_b)} {\cal P}_r = {\it I}, ~~ {\cal P}_r {\cal P}_s = \delta_{rs} {\cal P}_r, 
~~~ r,s = 0,1,2, \cdots \mbox{min} (n_a, n_b). 
\label{idon} 
\eea 
The orthogonality relation can be proven exactly like in the $SU(2)$ case (see (\ref{ido})) 
and the completeness relation, manifest in (\ref{suncgd}), is proved in appendix A.2. 

Note that the coupled states  are also eigenstates of $k_0$ and $k^2$ and carry the 
following Sp(2,R) quantum numbers: 
\bea 
k_0 ~{\cal P}_r~ |\{n_a\}\rangle \otimes |\{n_b\} \rangle   & = & 
\frac{1}{2}\left(n_a+n_b+N\right)~{\cal P}_r |\{n_a\}\rangle \otimes |\{n_b\} \rangle. \nonumber 
\label{sp2rnn} 
\\ \\ 
k^2 ~{\cal P}_r |\{n_a\}\rangle \otimes |\{n_b\} \rangle   &  = & 
q_0(1-q_0) ~{\cal P}_r  |\{n_a\}\rangle \otimes |\{n_b\} \rangle,  \nonumber  
\eea  
where, $q_0 \equiv \frac{1}{2}\left(n_a+n_b+N-2r\right).$ 
It reduces to its $SU(2)$ value for $N=2, n_a=2j_1, n_b=2j_2$.

\subsection{$SU(N) \times SU(N)$ irreducible Schwinger bosons} 
\label{sunisbs} 

Like in the section \ref{su2isbs} (also see \cite{su3isb,sunisb}), we define: 
\bea 
{A^\dagger}^\alpha \equiv {a^\dagger}^\alpha + F(\hat n_a,\hat n_b) k_+ {b}^\alpha,  ~~ B^\dagger_\alpha \equiv 
b^\dagger_\alpha + G(\hat n_a,\hat n_b) k_+ {a}_\alpha. 
\label{dpisbn}
\eea   
Note that by construction (\ref{dpisbn})  the $SU(N)_T$ transformation properties  of 
$A^{\dagger\alpha}$  and $B^\dagger_\alpha$  are exactly same as those of $a^{\dagger\alpha}$  
and $b^\dagger_\alpha$ respectively.  We now demand: 
\bea   
k_- ~A^{\dagger\alpha} ~{\cal P} |\{n^i_a\}\rangle \otimes |\{n^i_b\} \rangle = 0, ~~~ 
k_-~ B^\dagger_\alpha ~{\cal P} |\{n^i_a\}\rangle \otimes |\{n^i_b\} \rangle = 0. 
\label{conabn} 
\eea 
The above constraints can be solved in terms of the unknown functions $F(\hat n_a,\hat n_b)$ and $G(\hat n_a,\hat n_b)$: 
\bea 
F(\hat n_a,\hat n_b) = G(\hat n_a,\hat n_b)  = -\frac{1}{(\hat n_a+\hat n_b+N-2)}.
\label{confN}
\eea 
The first state of the Clebsch Gordon series for $SU(N)$$\times$$SU(N)$ as given in Figure \ref{sunf}  can be easily 
defined through the irreducible Schwinger bosons: 
\bea 
\Big|\{n_a\},\{n_b\};r=0 \Big\rangle & \equiv& 
{N}^{\{n_a\}}_{\{n_b\}}  
\frac{
\left(A^{\dagger 1}\right)^{n_a^1} 
\cdots 
\left(A^{\dagger N}\right)^{n_a^N} 
\left(B^\dagger_1\right)^{n_b^1} 
\cdots \left(B^\dagger_N\right)^{n_b^N}}
{\sqrt{\left(n_a^1\right)!\left(n_a^2\right)! \cdots \left(n_a^N\right)!\left(n_b^1\right)!\left(n_b^2\right)!
 \cdots \left(n_b^N\right)!}} ~\big|0\big\rangle \nonumber \\ \nonumber \\
&= &
{N}^{\{n_a\}}_{\{n_b\}} 
{\cal P} \frac{\left(a^{\dagger 1}\right)^{n_a^1} 
\cdots 
\left(a^{\dagger N}\right)^{n_a^N} 
\left(b^\dagger_1\right)^{n_b^1} 
\cdots \left(b^\dagger_N\right)^{n_b^N}}
{\sqrt{\left(n_a^1\right)!\left(n_a^2\right)! \cdots \left(n_a^N\right)!\left(n_b^1\right)!\left(n_b^2\right)!
 \cdots \left(n_b^N\right)!}} \big|0\big\rangle  \nonumber \\ \nonumber \\
&= & {N}^{\{n_a\}}_{\{n_b\}} {\cal P} 
\Big|\{n^i_a\}\Big\rangle \otimes \Big |\{n^i_b\}\Big\rangle 
\label{jm12n}
\eea 
The $r=0$ state in (\ref{jm12n}) is the first coupled representation on the 
right hand side of Figure \ref{sunf}. The  ${\cal N}^{\{n_a\}}_{\{n_b\}}$ are 
the normalization constants.  
Again the construction (\ref{jm12n}) is the simplest and direct generalization 
of  Schwinger boson construction  (\ref{dams}) and (\ref{dprn}) to $SU(N) \times SU(N)$ group. 
As an example we consider: 
$$\left(N^{\{n_a=1\}}_{\{n_b=1\}} \right)^{-1} |\{n_a=1\},\{n_b=1\}; r=0\rangle = 
A^{\dagger \alpha} B^{\dagger}_\beta |0\rangle =\left( a^{\dagger \alpha} b^{\dagger}_\beta - \frac{1}{N} 
\delta^\alpha_\beta k_+ \right) |0\rangle. $$
Thus the tracelesness or equivalently the symmetries of Young tableaues are manifestly present 
in the definition of $SU(N) \times SU(N)$ irreducible Schwinger bosons. 

\noindent Comparing (\ref{suncgc1}) at $r=0$ with  (\ref{jm12n}) we get: 
\bea 
{N}^{\{n_a\}}_{\{n_b\}} ~C^{~~~r=0}_{\{n_a\},\{n_b\}} =1. 
\label{norm2} 
\eea 
Hence, like in $SU(2)$ case (\ref{norm1}) the normalization factor 
${N}^{\{n_a\}}_{\{n_b\}}$ is just the inverse of the CG coefficient at $r=0$. This normalization 
can be calculated using (\ref{suncgc}) from section \ref{suncgcs}. 
As an example we consider SU(3) states (\ref{dprn}) with partitions: 
$n_a^1=1, n_a^3=0; n_b^1=1, n_b^2=1~,~ n_b^3=0$.
In (\ref{jm12n}) we replace the irreducible Schwinger bosons by their defining equation 
(\ref{dpisbn}) and (\ref{conf}) to get,
\bea 
&& \Big|\{n_a^1=1, n_a^2=0,n_a^3=0\},\{n_b^1=1,n_b^2=1,n_b^3=0\};r=0\Big\rangle 
 =  N^{1,0,0}_{1,1,0} ~~A_1^\dagger B^{\dagger 1} B^{\dagger 2} \big|0\big\rangle  \nonumber \\
& & =  N^{1,0,0}_{1,1,0}~~ \left[ \frac{3}{4} \left| 
\begin{array}{ccc} 1&0&0\\1&1&0 \end{array} 
\right\rangle -\frac{\sqrt{2}}{4} \left| \begin{array}{ccc} 0&1&0\\0&2&0
\end{array} \right\rangle 
- \frac{1}{4} \left| \begin{array}{ccc} 0&0&1\\0&1&1
\end{array} \right\rangle  \right]. 
\label{ole} 
\eea  
Therefore, explicit normalization of the above state gives: $N^{1,0,0}_{1,1,0} 
=\sqrt{\frac{4}{3}}$. On the other hand, this normalization can also be computed 
by using (\ref{norm2}) and the SU(3) Clebsch Gordan expression (\ref{suncgc}) obtained 
in the next section.
Putting the above values of occupation numbers and $r=0$ in (\ref{suncgc}) we get: 
$$\left(N^{1,0,0}_{1,1,0} \right)^{-1} =C^{r=0}_{\{n_a^1=1\},\{n_b^1=1,n_b^2=1\}}= 
\sqrt{\frac{2!}{4!3!3!}}(4!-3!)=\sqrt{\frac{3}{4}}.$$
Infact, at this stage we can cross check the other values of the $SU(N)$ Clebsch Gordan coefficients 
present in (\ref{ole}) with their values computed from the $SU(N)$ Clebsch Gordan expression (\ref{suncgc}) 
in the next section. The decomposition  (\ref{ole}) implies 
$$C^{~~r=0}_{\{010\}\{020\}} = -\sqrt{\frac{1}{6}}~~~ {\mathrm and} ~~~ 
C^{~~r=0}_{\{001\}\{011\}} = - \sqrt{\frac{1}{12}}.$$ 
As can be checked, these  are also the values obtained from (\ref{suncgc})  
after putting $N=3$, various occupation numbers and $r=0$. 
Thus the above simple state provides three self consistency checks on our procedure.  

The discussions in the previous section 
and Figure \ref{sunf} imply that  an arbitrary coupled state can be written as: 
\bea 
|\{n_a\}, \{n_b\};r\rangle =  
{\cal N}^{~r}_{n_a,n_b} \left(k_+\right)^{r} 
|\{n_a-r\},\{n_b-r\};r=0\rangle  
\label{fien}
\eea 
The normalization constants ${\cal N}^{~r}_{n_a,n_b}$ can be easily computed as $k_- 
|\{(n_a)\},\{(n _b)\};r=0\rangle =0$  and are given by: 
$${\cal N}^{~r}_{n_a,n_b} = \sqrt{\frac{(n_a+n_b+N-1)!}{r!(n_a+n_b+N+r-1)!}}.$$  
We again emphasize that except the invariant term  all the $SU(N)\times SU(N)$ 
coupled states in (\ref{fie}) are monomials of the irreducible Schwinger bosons. 
The present construction of coupled states is a straightforward generalization of the 
original construction to the decoupled $SU(2)$ angular momentum states (\ref{dams}). 

We note that the $SU(N) \times SU(N)$ irreducible Schwinger bosons satisfy: 
\bea 
\label{ccrn}
\left[A^\dagger_\alpha,A^\dagger_\beta\right] =0,~ \left[B^\dagger_\alpha,B^\dagger_\beta\right] =0,~
\left[A^{\dagger\alpha},B^\dagger_\beta\right] =0  \hspace{5cm}  \nonumber  \\ 
\left[A_\alpha,A^{\dagger\beta}\right] =  \delta^\beta{}_\alpha-\frac{1}{\hat n_a
+\hat n_b+N-1}b^\dagger_\alpha b^\beta +\frac{1}{(\hat n_a+\hat n_b+N-1)(\hat n_a+\hat n_b+N-2)}k_+a_\alpha b^\beta  \\
\left[B^\alpha,B^\dagger_\beta\right] = \delta^\alpha{}_\beta-\frac{1}{\hat n_a
+\hat n_b+N-1}a^{\dagger\alpha} a_\beta +\frac{1}{(\hat n_a+\hat n_b+N-1)(\hat n_a
+\hat n_b+N-2)}k_+a_\beta b^\alpha. \nonumber  
\eea 
These relations reduce to (\ref{ccr2}) for $N=2$. 

\subsection{The Projection operators and $SU(N)$ Clebsch Gordon Coefficients}
\label{suncgcs} 

We write (\ref{suncgd}) and (\ref{suncgc1}) as 
\bea
\label{suncg} 
\big|\{n_a\}\big\rangle \otimes \big|\{n_b\} \big\rangle 
= \sum_{r=0}^{n} \mathcal P_r 
\big|\{n_a\}\big\rangle \otimes \big|\{n_b\} \big\rangle 
\equiv \sum_{r=0}^{n} C^{~~~r}_{\{n_a\},\{n_b\}} \big|\{n_a-r\},\{n_b-r\};r\big\rangle
\eea
where, $n= \mbox{min}(n_a,n_b)$. Hence the Clebsch Gordon Coefficients can be computed as in the 
$SU(2)$ case:
\bea
C^{~~~r}_{\{n_a^i\},\{n_b^i\}} 
= \frac{~ \big\langle n_a,0,\cdots 0| \otimes \big\langle \bar n_b^1,\bar n_b^2,\cdots 
,\bar n_b^N\big| ~\mathcal P_r ~ \big|n_a^1, n_a^2,\cdots n_a^N \big\rangle \otimes \big|n_b^1,n_b^2,\cdots n_b^N
\big\rangle 
~~~}{\left[\big\langle n_a,0,\cdots ,0\big| \otimes \big\langle n_b^1,\bar n_b^2, \cdots , \bar n_b^N \big| 
~\mathcal P_r~ \big|  n_a,0,\cdots ,0 \big \rangle \otimes \big|\bar n_b^1,\bar n_b^2, \cdots ,\bar 
n_b^N\big\rangle\right]^{\frac{1}{2}}} 
\label{men}
\eea
In the above equation $\{\bar n_b^1,\cdots \bar n_b^N\}$ are the  values of the occupation numbers 
corresponding to the special choice $\{n_a^1=n^a,0,0,\cdots ,0\}$ so that the total magnetic quantum numbers 
on both sides of the projection operator remain unchanged\footnote{Note that 
the $SU(N)$ states $|n^1,n^2, \cdots ,n^N \rangle $ in (\ref{dprn}) can also be characterized by $SU(N)$ 
Casimir $n=n^1+n^2+\cdots +n^N$  along with the  ``$SU(N)$ magnetic quantum numbers" $\{h^i\} (i =1,2 \cdots (N-1))$ as:
\bea
h_a^1 = n_a^1-n_a^2, &&  h_b^1 = n_b^2-n_b^1  
\nonumber \\
h_a^2 = n_a^1+n_a^2-2n_a^3, &&   h_b^2 = 2n_b^3 - n_b^1 - n_b^2 \nonumber \\
&\vdots & \nonumber \\
h_a^{N-1}=n_a^1+n_a^2+\ldots +n_a^{N-1}-(N-1)n_a^N, &&  
h_b^{N-1}= (N-1)n_b^N  -n_b^1-n_b^2-\ldots -n_b^{N-1}. \nonumber 
\eea}: 
They are given by: 
\bea
 \bar n_b^1 = n_a-n_a^1+n_b^1 
 ~~~~~~{\mathrm and } ~~~~~\bar n_b^i = n_b^i-n_a^i ~~~~~, i=2,3,...,N\nonumber
\eea
As shown in appendix B.2, the matrix elements of $\mathcal P_r$ in (\ref{men}) can be easily computed to give,
\bea
\label{suncgc}
\hspace{1cm}C^{~~~r}_{\{n_a\},\{n_b\}}&=& \sqrt{ \frac{(n_a+n_b+N-2r-1)n_a^1!n_b^1!n_b^2!...n_b^N!}
{r!(n_a+n_b+N-r-1)!n_a^2!n_a^3!...n_a^N!\bar n_b^2!...\bar n_b^N! } } \nonumber \\ &&
\sqrt{ \frac{(n_a-r)! (\bar n_b^1-r)! (n_b-\bar n_b^1+N-2)!}{(n_b+N-r-2)!(n_a+n_b-\bar n_b^1+N-r-2)!} }\nonumber \\ &&
\hspace{-2cm} \sum_{q}^{\mbox{min}(n_a-r, n_b-r)} \frac{(-1)^q}{q!} \frac{(q+r)!(n_a+n_b+N-2-2r-q)!}
{(n_a-q-r)!(\bar n_b^1-q-r)!(n_a^1-n_a+q+r)!}
\eea

Note  that this $SU(N)$ Clebsh Gordon series reduces to the $SU(2)$ Clebsch Gordon series (\ref{ffcg}) 
for $N=2$. This can be checked by identifying  
$(b^\dagger_2, b^\dagger_1)$ of $SU(N)$ with  $(b^\dagger_1, -b^\dagger_2)$ of $SU(2)$ respectively 
 so that $a^\dagger\cdot b^\dagger$ ($SU(N)$ invariant) $ \rightarrow a^\dagger\cdot \tilde b^\dagger$ 
($SU(2)$ invariant) 
and putting: 
\bea
\begin{array}{ccc} n_a^1= j_1+m_1 ~~&  n_b^1= j_2-m_2 ~~ &   \bar n_b^1= j_2-(m-j_1)  \\ \nonumber  \\ 
                    n_a^2= j_1-m_1 ~~&  n_b^2= j_2+m_2 ~~ &   \bar n_b^2= j_2+(m-j_1). \end{array}  
\eea 
The $SU(2)$ Casimirs in (\ref{suncgc}) are: $n_a= n_a^1+n_a^2 = 2j_1,~~ n_b= n_b^1+n_b^2 = 
\bar n_b^1+\bar n_b^2 =2j_2$ and $r= j_1+j_2-j$.

\section{Summary and discussions}

In this work, we have investigated the role of $SU(N) \times SU(N)$ invariant groups in the 
Clebsch Gordan decomposition of direct product of two $SU(N)$ irreducible representations. The 
techniques were completely based on the invariant groups and their algebras 
enabling us to handle all $SU(N)$ within a single framework.  It was crucial to use Schwinger construction
to get all possible invariants. 
The invariant group 
generators were used to construct  projection operators to get all possible coupled irreducible representations.  
The $SU(N)$ Clebsch Gordan coefficients were computed as matrix elements of these projection operators. 
Using the invariant algebra we also constructed $SU(N) \times SU(N)$ 
irreducible Schwinger bosons which directly creates the coupled irreducible states.  
Note that  in the case of  $SU(N)$ ($ N \ge 3$) we only considered direct product of 
$N$ and $N^*$ representations leading to Sp(2,R) invariant algebras. In fact the analysis of section \ref{suninvs} 
is  also valid for any two $SU(N)$ fundamental conjugate representations of dimensions ${}^{{}^N} C_r$ each. For 
simplicity we had chosen $r =1$. It will be interesting to extend these techniques to direct product of two 
arbitrary $SU(N)$ irreducible representations.  The invariant group involved will then be much larger. 
All possible projection operators and the irreducible Schwinger bosons will 
again depend on the invariant operators or generators of the invariant group. 
The  work in this direction is in progress 
and will be reported elsewhere.   

\newpage 

\appendix

\section{The projection operators}

In this appendix we construct and prove the completeness property of $SU(N) \times SU(N)$ projection 
operators. 

\subsection{SU(2)$\times$SU(2)}

We start with the construction of projection 
operator associated with symmetrization: 
\bea
\mathcal P\left(|j_1,m_1\rangle\otimes |j_2,m_2\rangle\right)= C_{j_1,m_1;j_2,m_2}^{~~j, m} 
|j_1,j_2,j=j_1+j_2,m=m_1+m_2\rangle
\label{aa1} 
\eea
We note that the $SU(2)\times SU(2)$ transformation property as well as  the total number of $a^\dagger$'s 
($=2j_1=n_a$) and $b^\dagger$'s ($=2j_2=n_b$) are same for the decoupled and coupled states on the left and 
right hand side of (\ref{aa1}). Therefore the projection operator is of the form:
\bea
\mathcal P= 
\sum_{q=0}^{\mbox{min}\{ n_a,n_b\}} l_q(\hat n_a,\hat n_b) k_+^qk_-^q
\label{aa22} 
\eea
where, $k_\pm$ are the $SU(2)$ invariant Sp(2,R) operators defined in (\ref{sp2r1o}). The unknown coefficients 
$l_q(\hat n_a, \hat n_b)$ can be easily fixed by demanding that the projected state is completely symmetric 
in all the $SU(2)$ indices and therefore should be annihilated by $k_-0$, i.e,
\bea
\label{p}
k_- ~\mathcal P \mathcal~ |j_1,m_1\rangle \otimes |j_2,m_2\rangle = 0. 
\eea
After using $k_-k_+^q  = \left[k_-,k_+^q\right] =q(\hat n_a+\hat n_b-q+3)k_+^{q-1}$ we get the 
recurrence relation: 
$$l_{q+1}(\hat n_a, \hat n_b) = -\frac{1}{(q+1)(\hat n_a+ \hat n_b-q)}l_q(\hat n_a,\hat n_b),$$  
leading to: 
\bea 
l_q (\hat n_a,\hat n_b)=\frac{(-1)^q}{q!}\frac{(\hat n_a+\hat n_b-q)!}{(\hat n_a+\hat n_b)!}
\label{aa2}
\eea 
Note that  $l_{0}(\hat n_a, \hat n_b)=1$ implying the projection operator ${\cal P}^2 ={\cal P}$.

To compute  the $\mathcal P_r$ normalization coefficients 
in (\ref{conp}) we use the  required ${\cal P}^2_r  ={\cal P}_r$ property:   
\bea 
{\cal P}_r ~ {\cal P}_r  &= & N_r^2 ~\{(k_+)^r  ~{\cal P} ~ (k_-)^r \}~\{(k_+)^r  ~{\cal P} ~ (k_-)^r\} 
 =   N_r^2 ~(k_+)^r ~ {\cal P} ~\left[(k_-)^r,(k_+)^r\right] ~{\cal P}
~(k_-)^r \nonumber \\
&= & \left(N_r r!\frac{(n_a+ n_b-r+1)!}{(n_a+ n_b-2r+1)!}\right)  
\underbrace{N_r~(k_+)^r  {\cal P}~ (k_-)^r}_{={\cal P}_r}   
=  {\mathcal P}_r. 
\label{ido1} 
\eea 
In (\ref{ido1}) we have used $k_-{\cal P} =0$ to replace $k_-^rk_+^r$ by the commutator 
$\left[k_-^r,k_+^r\right] \left(= r!\frac{(\hat n_a+ \hat n_b+r+1)!}{(n_a+ n_b+1)!}\right)$ and 
replaced the number operators $\hat n_a,\hat n_b$ by their eigenvalues 
$ n_a =2j_1, n_b=2j_2$ at the end. The above eqn. gives: 
\bea 
N_r = \frac{(n_a+n_b-2r+1)!}{r!(n_a+n_b-r+1)!} 
\label{connpa}
\eea 
One can easily check that ${\cal P}_r {\cal P}_{s \neq r}  = 0$ as $k_-{\cal P} =0 (r > s)$ and 
${\cal P} k_+ = 0 (r < s)$ leading to orthonormal irreducible Hilbert spaces characterized by 
the net angular momentum quantum numbers. 

\noindent We now prove the completeness  property of the projection operators ${\cal P}_r$. 
We start with: 
\bea
\sum_{r} 
\mathcal P_r | j_1,m_1 \rangle \otimes |j_2,m_2\rangle  
& = &\sum_{r}~ {\cal P}_r ~ \sum_{j=|j_1-j_2|}^{j_1+j_2} C_{j_1,m_1;j_2,m_2}^{~~j,m} |j_1,j_2;j,m\rangle 
\nonumber \\
&=& \sum_{r} {\cal P}_r ~\sum_{j=|j_1-j_2|}^{j_1+j_2} C_{j_1,m_1;j_2,m_2}^{~~j,m} {N}_{j,m} 
\left(J^-\right)^{j-m} |j_1,j_2;j,m=j\rangle \nonumber \\ 
&=& 
\sum_{j=|j_1-j_2|}^{j_1+j_2} C_{j_1,m_1;j_2,m_2}^{~~j,m} {N}_{j,m} 
\left(J^-\right)^{j-m} \underbrace{\sum_{r} {\cal P}_r~ |j_1,j_2;j,m=j\rangle}_{={\cal P}_0 |j_1,j_2;j,m=j\rangle
= |j_1,j_2;j,m=j\rangle}   \nonumber \\ 
&=& | j_1,m_1 \rangle \otimes |j_2,m_2\rangle. 
\label{crr}
\eea
In (\ref{crr}) the total lowering operator is defined as $J^- \equiv J_a^- + J_b^- (= a^\dagger_1a_2 + 
b^\dagger_1b_2)$ and $N_{j,m} (= \sqrt{\frac{(j+m)!}{(2j)!(j-m)!}} $ are the corresponding constants. We have also used the fact 
that the projection operators ${\cal P}_r$ commute with $J^-$ 
and satisfy orthonormality condition (\ref{ido}). 


\subsection{SU(N)$\times$SU(N)}

We can exactly follow the $SU(2)$ techniques of the previous section and use the relation 
$\left[k_-,k_+^q\right] =q(\hat n_a+\hat n_b+N+1-q)k_+^{q-1}$ to obtain the $SU(N) \times SU(N)$ 
projection operator (\ref{sunpo1}): 
\[L_q (n_a,n_b)=\frac{(-1)^q}{q!}\frac{(n_a+n_b+N-2-q)!}{(n_a+n_b+N-2)!} \]. 
Similarly, as in $SU(2)$ case (\ref{ido}):  
\bea 
{\cal P}_r ~ {\cal P}_r &= & N_r^2 \{k_+^r  {\cal P} k_-^r \}\{k_+^r  {\cal P}  k_-^r\} 
 =  N_r^2 k_+^r  {\cal P} ~ \left[k_-^r,k_+^r\right] ~ {\cal P}~k_-^r 
\nonumber \\
& = &   N_r r!\frac{(n_a+n_b+N-r-1)!}{(n_a+n_b+N-2r-1)!}{\cal P}_r \equiv {\cal P}_r.  
\label{idonn} 
\eea 
Above we have used the relation $\left[k_-^r,k_+^r\right] =  
\frac{r!(\hat n_a+\hat n_b+N+r-1)!}{(\hat n_a+\hat n_b+N-1)!}.$ 
We thus get: 
\bea 
N_r = \frac{(n_a+n_b+N-2r-1)!}{r!(n_a+n_b+N-r-1)!} 
\label{conpan}
\eea 
As in $SU(2)$ case the different projected or irreducible spaces are orthonormal: ${\cal P}_r {\cal P}_{s \neq r} 
= 0$ as $k_-{\cal P} =0 ( r >  s)$ and ${\cal P} k_+ = 0 
(r < s).$  The completeness property of the $SU(N)$ projection operators also follows exactly as in the $SU(2)$ case.
 
\section{Matrix elements of Projection operators}

In this appendix we compute the matrix elements of projection operators in (\ref{cgce2}) and (\ref{men}) 
to get the $SU(2)$ and $SU(N)$ Clebsch Gordan coefficients. 

\subsection{SU(2)$\times$SU(2)}

The numerator in (\ref{cgce2}) is: 
\bea
&& \hspace{-1.0cm} \langle j_1,j_1,j_2,m -j_1| {\cal P}_r|j_1,m_1,j_2,m_2\rangle   =  
N_r \left\langle \begin{array}{cc} 2j_1 & 0 \\ j_2+m-j_1 & j_1+j_2-m \end{array} 
\right\vert k_+^r {\cal P} k_-^r  \left\vert  \begin{array}{cc} j_1 +m_1& j_1-m_1 \\ j_2+m_2  & j_2-m_2  \end{array} 
\right\rangle  \nonumber \\ 
& = & \hspace{-0.4cm} N_r \sum_{q} l_q(2j_1-r,2j_2-r) 
\underbrace{\left\langle \begin{array}{cc} 2j_1 & 0 \\ j_2+m-j_1 & j_1+j_2-m \end{array} 
\right\vert k_+^{q+r}~ k_-^{q+r} \left\vert  \begin{array}{cc} j_1 +m_1& j_1-m_1 \\ j_2+m_2  & j_2-m_2  \end{array} 
\right\rangle}_{\equiv K(j_1,m_1,j_2,m_2,q,r)}  \nonumber \\ 
& = & \hspace{-0.4cm} N_r \sum_{q} ~l_q(2j_1-r,2j_2-r)~  K(j_1,m_1,j_2,m_2,q,r).   
\label{me2}
\eea 
In the first step we have written the decoupled angular momentum states in terms of the occupation number 
basis.  In the second step we have substituted the expansion (\ref{aa22}) of ${\cal P}$  with 
$n_a=2j_1-r~,~ n_b=2j_2-r$ for the coefficient $l_q$ in (\ref{aa2}).
Note that the matrix elements K  can be easily computed as both  
$k_+^{q+r}$ and $k_-^{q+r}$  in (\ref{me2}) 
can be replaced by monomials of harmonic oscillator creation and annihilation operators respectively:   
\bea 
k_+^{q+r}  \rightarrow   \left(a^\dagger_1~b^\dagger_2 \right)^{q+r}, ~~~~~~~~~ 
k_-^{q+r} \rightarrow   (-1)^{q+r-s} ~{}^{{}^{q+r}}C_s ~\left(a_1b_2\right)^{s} 
\left(a_2b_1\right)^{q+r-s}.  \nonumber 
\eea 
Above $s = q+r+m_1-j_1$.  Substituting these monomials in (\ref{me2}) 
leads to: 
\bea  
K
&=&  \frac{(-1)^{q+r-s}(q+r)!}{s!(q+r-s)!(j_1+m_1-s)! (j_2-m_2-s)!(j_1-m_1-q-r+s)!(j_2+m_2-q-r+s)!} \nonumber 
\\ \nonumber \\ && \times \sqrt{ (2j_1)! (j_2+m-j_1)!
(j_2-m+j_1)! (j_1+m_1)! (j_1-m_1)! (j_2+m_2)! (j_2-m_2)!}.   
\label{kk}
\eea 
Substituting $N_r$ from (\ref{connpa}), $l_q(2j_1-r,2j_2-r)$ from (\ref{aa2}) and K from above with $s=q+r+m_1-j_1$,
the matrix element (\ref{me2}) takes the form: 
\bea
\langle j_1,j_1,j_2,m -j_1| {\cal P}_r|j_1,m_1,j_2,m_2\rangle  &=& \frac{(2j_1+2j_2-2r+1)!}{r!(2j_1+2j_2-r+1)!
(2j_1+2j_2-2r)!} \nonumber \\  \nonumber  \\&&
\hspace{-1cm}\times\sqrt{ \frac{(2j_1)! 
(j_2-m+j_1)! (j_1+m_1)! (j_2+m_2)! (j_2-m_2)!}{(j_2+m-j_1)! (j_1-m_1)! }}\nonumber \\ &&
\hspace{-3.1cm}\sum_{q=0}^{\mbox{min}(2j_1-r, 2j_2-r)} 
\hspace{-0.8cm}\frac{(-1)^{q+j_1-m_1}(q+r)!(2j_1+2j_2-2r-q)!}{(q+r+m_1-j_1)!(j_1+j_2-m-q-r)!(2j_1-q-r)!}
\eea
Putting $r=j_1+j_2-j$ in the above equation we get:
\bea
\label{num2}
\langle j_1,j_1,j_2,m -j_1| {\cal P}_j|j_1,m_1,j_2,m_2\rangle &=& 
\Bigg[\frac{(2j+1)!}{(2j)!(j_1+j_2-j)!(j_1+j_2+j+1)!}\nonumber \\ \nonumber \\
&& \hspace{-2cm} \sqrt{ \frac{(2j_1)!(j_1+j_2-m)!(j_1+m_1)!(j_2+m_2)!(j_2-m_2)!}{(j_1-m_1)!(j_2-j_1+m)!}}~\Bigg] 
\nonumber \\ \nonumber \\
&&\hspace{-7cm} \sum_{q=0}^{\mbox{min}(j_1-j_2+j, j_2-j_1+j)} \frac{(-1)^{q+j_1-m_1}}{(q)!} \frac{(j_1+j_2-j+q)!(2j-q)!}{(j_2-j+q+m_1)!(j_1-j_2+j-q)!(j-m-q)!} 
\eea
For the denominator of (\ref{cgce2}), we substitute $m_1=j_1$ and $m_2=m-j_1$ in (\ref{num2}) to obtain,
\bea
&&\hspace{2.9cm} \langle j_1, j_1, j_2, m-j_1| \mathcal P_r|j_1, j_1, j_2 ,m-j_1 \rangle 
\nonumber \\ \nonumber \\ 
&& \hspace{-0.6cm} = \frac{(2j+1)!(2j_1)!(j_2+j_1-m)!}{(2j)!(j_1+j_2-j)!(j_1+j_2+j+1)!}
~\sum_{q=0}^{q_{\textrm max}} ~ 
\frac{(-1)^q}{q!} \frac{(2j-q)!}{(j_1-j_2+j-q)!(j-m-q)!}.
\label{den2}
\eea
In (\ref{den2}) the upper limit on the sum over q is 
$q_{{\max}} \equiv  {min}(2j_1-r,2j_2-r) = {min} (j_1-j_2+j, j_2-j_1+j).$ 
This above series in q  is summed using the formula:
\bea
\label{densum}
\sum_{q=0} \frac{(-1)^q}{q!} \frac{(C-q)!}{(A-q)!(B-q)!}= \frac{C!}{A!B!}\times\frac{(C-A)!(C-B)!}{C!(C-A-B)!}
\eea
Finally, the denominator in (\ref{cgce2}) is: 
\bea
\label{denf2}
&&\hspace{3cm} \sqrt{\langle j_1, j_1, j_2, m-j_1| \mathcal P_r|j_1, j_1, j_2 ,m-j_1 \rangle } \nonumber 
\\&&= \sqrt{ \frac{(j+m)!(j_2-j_1+j)!(2j+1)!(2j_1)!(j_2+j_1-m)!}{(2j)!(j_1+j_2-j)!(j_1+j_2+j+1)!
(j-m)!(j_2-j_1+m)!(j_1-j_2+j)!} }
\eea
The final expression of the Clebsch Gordon coefficient in (\ref{ffcg}) is now obtained by dividing (\ref{num2}) by 
(\ref{denf2}).

\subsection{SU(N)$\times$SU(N)}

Similarly the matrix element in the numerator of $SU(N)$ Clebsch Gordon coefficient expression (\ref{men}) is:
\bea
&&\hspace{3cm} \Big\langle \{n_a^1=n_a\},\{\bar n_b\}\Big| \mathcal P_r \mathcal P_r 
\Big|\{n_a\},\{n_b\}\Big\rangle  \nonumber \\&=& N_r \sum_{q} l_q(n_a-r,n_b-r)\underbrace{\left\langle 
\begin{array}{cccc} n_a & 0& \ldots & 0 \\ \bar n_b^1 & \bar n_b^2 &\ldots &\bar n_b^N \end{array}\right\vert 
k_+^{q+r}k_-^{q+r}\left\vert \begin{array}{cccc} n_a^1 & n_a^2& \ldots & n_a^N \\ n_b^1 &  n_b^2 &\ldots & 
n_b^N \end{array} \right\rangle }_{K\left(\{n_a\}, \{n_b\}, q, r\right)}\nonumber \\&& \hspace{2cm} = 
N_r \sum_{q} l_q(n_a-r,n_b-r) K\left(\{n_a\}, \{n_b\}, q, r\right)
\label{sunxx} 
\eea
The matrix element $ K\left(\{n_a\}, \{n_b\}, q, r\right) $ are calculated in the same way as in the  $SU(2)$ case.
In the computation of $K\left(\{n_a\}, \{n_b\}, q, r\right)$ in (\ref{sunxx})  $k_+^{q+r}$ 
and $k_-^{q+r}$ can be replaced by  the following monomials of Schwinger bosons:
\bea
k_+^{q+r}\rightarrow (a^\dagger_1 b^{\dagger 1})^{q+r}, ~~~~~~~~ k_-^{q+r}\rightarrow \frac{(q+r)!}{\beta_1!\ldots 
\beta_N!} (a^1 b_1)^{\beta_1} (a^2 b_2)^{\beta_2}\ldots(a^N b_N)^{\beta_N}
\eea
Equating the occupation numbers in the matrix element in (\ref{sunxx}) we get: 
$$\beta_1= q+r+n_a^1-n_a,~~\beta_2= n_a^2,~~\beta_3= n_a^3, ~~~~\ldots,\beta_N= n_a^N,$$ 
leading to: 
\bea
K\left(\{n_a\}, \{n_b\}, q, r\right)&=& 
\sqrt{ n_a!\bar n_b^1!\bar n_b^2!\ldots\bar n_b^N!n_a^1!n_a^2!\ldots n_a^N!  n_b^1!n_b^2!\ldots  n_b^N!  }
\nonumber \\&& 
\times \frac{(q+r)!}{(n_a^1-n_a+q+r)!n_a^2!\ldots n_a^N!}\nonumber \\
&& \times \frac{1}{(n_a-q-r)!(\bar n_b^1-q-r)!\bar n_b^2!\ldots\bar n_b^N!}
\eea
Now substituting the values of $N_r$ and $l_q(n_a-r,n_b-r)$ from A.2 and the matrix element $K$ from above we finally get the numerator of (\ref{men}) as:
\bea
\Big\langle \{n_a^1=n_a\},\{\bar n_b\}\Big| \mathcal P_r  \Big|\{n_a\},\{n_b\}\Big\rangle&=&
\frac{(n_a+n_b+N-2r-1)}{r!(n_a+n_b+N-r-1)!}\sqrt{ \frac{n_a!\bar n_b^1!n_a^1! n_b^1!\ldots  n_b^N!}{ n_a^2!\ldots n_a^N!\bar n_b^2!\ldots\bar n_b^N! } }\nonumber \\&&\sum_{q} \frac{(-1)^q (q+r)!(n_a+n_b+N-2-2r-q)! }{q!(n_a-q-r)!(\bar n_b^1-q-r)!(n_a^1-n_a+q+r)!}
\label{numerator}
\eea
Like in  $SU(2)$ case, the denominator of (\ref{men}) is the square-root of the numerator with 
$n_a^1=n_a$, $n_a^2=n_a^3=\cdots =n_a^N=0$ and $n_b^i=\bar n_b^i$, $\forall i$. 
The final expression for the denominator in (\ref{men}) is:
\bea
&&\Big\langle \{n_a^1=n_a\},\{\bar n_b\}\Big| \mathcal P_r \Big|\{n_a^1=n_a\},\{\bar n_b\}\Big\rangle \nonumber \\
&=&
\frac{(n_a+n_b+N-2r-1)n_a!\bar n_b^1!}{r!(n_a+n_b+N-r-1)!}\sum_{q} 
\frac{(-1)^q (n_a+n_b+N-2-2r-q)! }{q!(n_a-q-r)!(\bar n_b^1-q-r)!} \nonumber \\&& \nonumber \\&=&
\frac{(n_a+n_b+N-2r-1)n_a!\bar n_b^1!(n_b+N-r-2)!(n_a-\bar n_b^1+n_b+N-2-r)!}{r!(n_a+n_b+N-r-1)!(n_a-r)!
(\bar n_b^1-r)!(n_b-\bar n_b^1+N-2)!}. 
\label{denominator1}
\eea
In (\ref{denominator1}) the last sum has been performed  using (\ref{densum}) again.
Finally, the $SU(N)$ Clebsch Gordon coefficient expansion (\ref{suncgc}) is obtained by dividing (\ref{numerator}) 
with square root of (\ref{denominator1}).

\end{document}